# How to Directly Image a Habitable Planet Around Alpha Centauri with a ~30-45cm Space Telescope

Ruslan Belikov*[a], Eduardo Bendek[a], Sandrine Thomas[b], Jared Males[c], Julien Lozi[d],
and the ACESat team

[a]NASA Ames Research Center, Moffett Field, CA 94035, USA;
[b]Large Synoptic Survey Telescope, 950 N Cherry AVe, Tucson AZ 85719 USA
[c]Steward Observatory, 933 North Cherry Avenue, Tucson, AZ 85721, USA;
[d]Subaru Telescope, NAOJ, 650 North A'ohoku Place, Hilo, HI 96720, USA

## ABSTRACT

Several mission concepts are being studied to directly image planets around nearby stars. It is commonly thought that directly imaging a potentially habitable exoplanet around a Sun-like star requires space telescopes with apertures of at least 1m. A notable exception to this is Alpha Centauri (A and B), which is an extreme outlier among FGKM stars in terms of apparent habitable zone size: the habitable zones are ~3x wider in apparent size than around any other FGKM star. This enables a ~30-45cm visible light space telescope equipped with a modern high performance coronagraph or starshade to resolve the habitable zone at high contrast and directly image any potentially habitable planet that may exist in the system. We presents a brief analysis of the astrophysical and technical challenges involved with direct imaging of Alpha Centauri with a small telescope and describe two new technologies that address some of the key technical challenges. In particular, the raw contrast requirements for such an instrument can be relaxed to 1e-8 if the mission spends 2 years collecting tens of thousands of images on the same target, enabling a factor of 500-1000 speckle suppression in post processing using a new technique called Orbital Difference Imaging (ODI). The raw light leak from both stars is controllable with a special wavefront control algorithm known as Multi-Star Wavefront Control (MSWC), which independently suppresses diffraction and aberrations from both stars using independent modes on the deformable mirror. We also show an example of a small coronagraphic mission concept to take advantage of this opportunity.

## 1. MOTIVATION: UNIQUE OPPORUNITY ENABLED BY ALPHA CENTAURI

The search for another Earth-like planet (exo-Earth) and extraterrestrial life is one of the most fundamental, grand, and noble pursuits not just in astronomy, but in all of science. The discovery of a true exo-Earth is likely to be hailed as a major milestone of our civilization, on par with the landings on the Moon and humanity's first steps in space. Several mission concepts are currently being studied to directly image planets around nearby stars, with varying capabilities and sensitivities. It is commonly thought that directly imaging a potentially habitable planet around a Sun-like star requires telescope apertures of at least 1m, costing at least $1B, and launching no earlier than the mid-2020s and more likely the 2030s. This conventional wisdom is probably correct for all stars, with one significant exception: Alpha Centauri. As we will show in this paper, a sufficiently powerful 30-45cm high contrast space telescope is in principle sufficient to directly image any potentially habitable (and many other) planets around Alpha Centauri A and B.

Figure 1 shows the current landscape of the field of exoplanet direct imaging, with different color shaded regions representing the capabilities of current, past, and future direct imaging instruments in visible and near infrared light. The upper right hand corner represents the capability of previous-generation coronagraphs (many of which remain competitive and continue to produce significant results). The dots in the upper right hand corner represent a sampling of planets or planetary mass objects that have been directly imaged to date. All of these are young, hot, self-luminous, large planets or objects orbiting very far from their star, at contrasts of roughly $10^{-6}$ to $10^{-4}$. Note that the capability of directly imaging with much deeper contrasts (better than $10^{-9}$) has already existed for many years as evidenced by Fomalhaut b (for large separations such as 10"). Therefore, contrast alone is often not a meaningful performance metric unless referred to a particular separation angle. The current generation of "Extreme Adaptive Optics" coronagraphs such as GPI, SPHERE, P1640, SCExAO push the performance boundary to the lower left into more aggressive combination of contrasts and separation angles. These instruments are expected to find many more young giant self-luminous planets, and perhaps some mature planets shining by reflected light as well. However, they will fall short of detecting potentially habitable planets.

*Ruslan.Belikov@nasa.gov; phone 1 650 604 0833

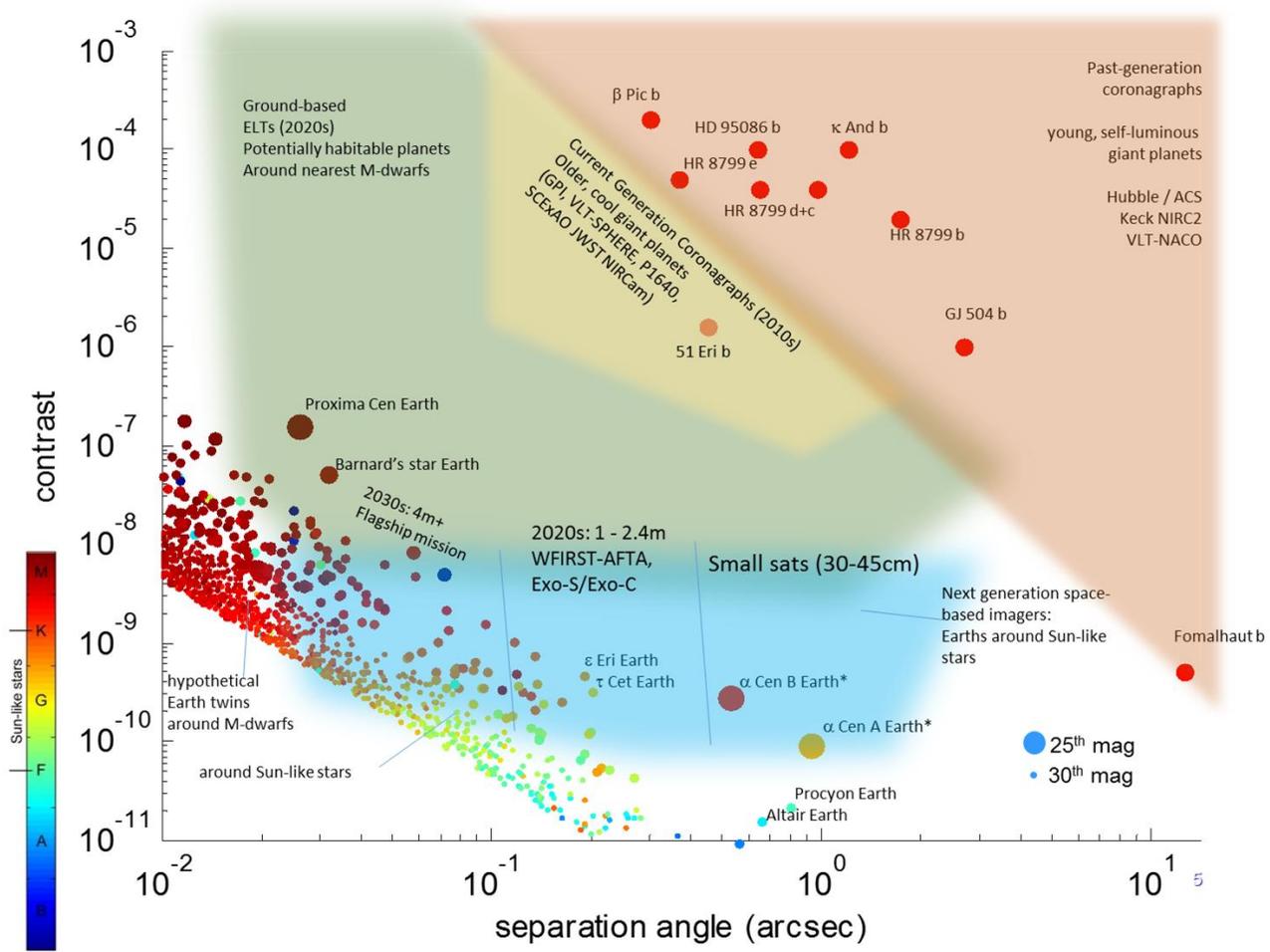

Figure 1. Capabilities of present and future high contrast imaging instruments and types of planets they can image. A sampling of already imaged planets appears in the upper right corner. The circles in the bottom left corner show where an Earth twin in the HZ would be around every nearby star out to 20pc (color of the circle represents the color or type of the star and area of the circle is proportional to apparent planet brightness). Ground-based ELTs and next generation space-based imagers have the potential to directly image and take the spectra of potentially habitable worlds around M-dwarfs, but not around sun-like stars. Next generation of space-based imagers will have the capability to directly image Earth twins around Sun-like stars. Small satellites can be launched as early as late this decade capable of detecting Earth-like planets around the nearest star Alpha Centauri (A and B). In the 2020s, the WFIRST-AFTA, or one of Exo-C or Exo-S may directly image a few potentially habitable worlds in the 2020s. Ultimately, a large space based coronagraphic telescope like LUVOIR or HabEx will have the ability to search hundreds of stars for potentially habitable worlds and take spectra with enough resolution to detect biomarkers like oxygen and water.

The dots on the lower left corner represent hypothetical Earth twins at quadrature around every nearby star out to 20pc. The color represents the star type (see colorbar on the bottom left corner) and dot size represents the magnitude of the planet, which ranges from about $25^{th}$ for the nearest stars to about $31^{st}$ magnitude for the farthest stars. (The diagonal cutoff edge of the scatter points represents Earth twins equidistant from our Earth at 20pc, and lines parallel to it represent planets that are equidistant from us). Ground-based Extremely Large Telescopes expected to start operations in the mid-2020s may be able to reach ~$10^{-8}$ contrasts at separations approaching 10-20mas (if equipped with powerful coronagraphs). This will enable them to detect potentially habitable planets around the nearest M-dwarfs. However, potentially habitable planets around Sun-like stars will appear at contrasts of $10^{-9}$ and deeper, and achieving such contrasts may not be possible from the ground (even out at larger separations such as ~1") because the number of atmospheric modes goes up with aperture, effectively cancelling out the benefits due to higher number of photons collected.

Thus, in order to achieve the $10^{10}$ contrasts to directly image potentially habitable planets around Sun-like stars, high contrast space telescopes (blue region in Figure 1) appear to be necessary. High contrast space telescopes in the 1 - 2.4m size range are realistic for launch in the 2020 decade, such as NASA's planned WFIRST-AFTA mission [1], as well as the Exo-C and Exo-S concept studies [2, 3]. Such telescopes can in principle access the habitable zones of 10s of Sun-like stars, depending on coronagraphic inner working angle. Indeed, both Exo-C and Exo-S missions have been designed to be capable of directly imaging potentially habitable worlds. WFIRST-AFTA has not, but some simulations show that it may be able to directly image such worlds [1]. The 2030s may prove to be an exciting era of an even larger high contrast space telescope of 4m or higher aperture [4,5], capable of surveying hundreds of stars for potentially habitable planets and taking spectra with high enough resolution to determine detailed atmospheric compositions and truly search for signs of life.

These missions will surely make great leaps in the exploration of exoplanets, habitable or not, around many nearby stars. However, they are large, expensive, and still about a decade or more away. Alpha Centauri, the nearest star system to us, may offer an opportunity to directly image a planetary system, including the habitable zone, with a much smaller telescope at much lower cost and potentially late this decade or early next. This is because Alpha Centauri is not merely the closest star system to the Sun, but happens to be an unusual outlier among FGKM stars in terms of its apparent habitable zone size. Specifically, α Cen A & B habitable zones span 0.4-1.6" in stellocentric angle (see Table 1), ~3x wider than around any other FGKM star (where we used the "optimistic" habitable zones in [6]). This angle is so large that it actually falls outside the *outer* working angle of many large high contrast telescopes. In theory this enables a visible light telescope as small as 30cm, equipped with a modern high performance coronagraph, to resolve the habitable zone at high contrast and directly image any potentially habitable planet that may exist in the system. Due to the extreme apparent brightness of α Cen AB, exposure times can be as short as minutes with ideal components, or days with realistic ones. This makes it possible to do color photometry on potentially habitable planets sufficient to differentiate Venus-like, Earth-like, and Mars-like planets from each other and constrain the presence of Earth-pressure atmosphere through Rayleigh scattering.

*Table 1: Habitable zones for α Cen AB.*

| Orbit Range | Sun at α Cen Distance | α Cen A | | | α Cen B | | |
|---|---|---|---|---|---|---|---|
| | Semi-major axis (AU) | Period (years) | angular sep (as) | angular sep (l/D for 0.45m @550nm) | Period (years) | angular sep (as) | angular sep (l/D for 0.45m @550nm) |
| Earth twin | 1 | 1.3 | 0.92" | 3.6 | 0.62 | 0.54" | 2 |
| Habitable zone | 0.75-1.77 | 0.85-3.07 | 0.7-1.6" | 2.7-6.5 | 0.41-1.47 | 0.4-0.95" | 1.6-3.8 |

As Figure 1 shows, the α Cen system is easier than the next easiest target by almost any metric. In particular, the next nearest star after α Cen is Barnard's star, which is only 1.4 times farther, but is thousands of times dimmer than α Cen and has a habitable zone only 0.03" wide, requiring at least a 4m aperture telescope to even resolve it. It is also an M-dwarf which may not be ideal for habitable planets. The next closest star brighter than an M-dwarf (ε Eri) is 2.4 times as far, again requiring a much larger telescope to resolve its habitable zone. It is also known to have a thick disk that may interfere with detection of small planets. There are a handful of nearby stars whose habitable zone are resolvable with telescope as small as ~30cm (such as Sirius and Procyon), but these stars are much brighter (earlier type) than the Sun, making the star-planet contrast an order of magnitude more challenging which translates to the need for a much larger telescope (all else being the same). Such stars may also be too short-lived for life, especially advanced life, to develop on their planets.

## 2. ASTROPHYSICAL CONSIDERATIONS AND NOISE

### 2.1 Dynamical Stability

An important consideration for the existence of habitable terrestrial planets is their dynamical stability on long timescales. There have been a number of studies modeling planet formation and stability in the α Cen system [7-12, etc.] We adopt the stability limits of [7] who determined that the orbital semi-major axis within which planets are stable is 2.7789 +/- 0.6513 AU and 2.4922 +/- 0.7076 AU for α Cen A and B, respectively (see Figure 2). Planet formation models suggest that planetary orbits are likely to be prograde and near the plane of the binary, as shown in Figure 2. These stability limits mean that with a modest OWA requirement, a small space telescope can image the entire planetary system out from some inner working angle.

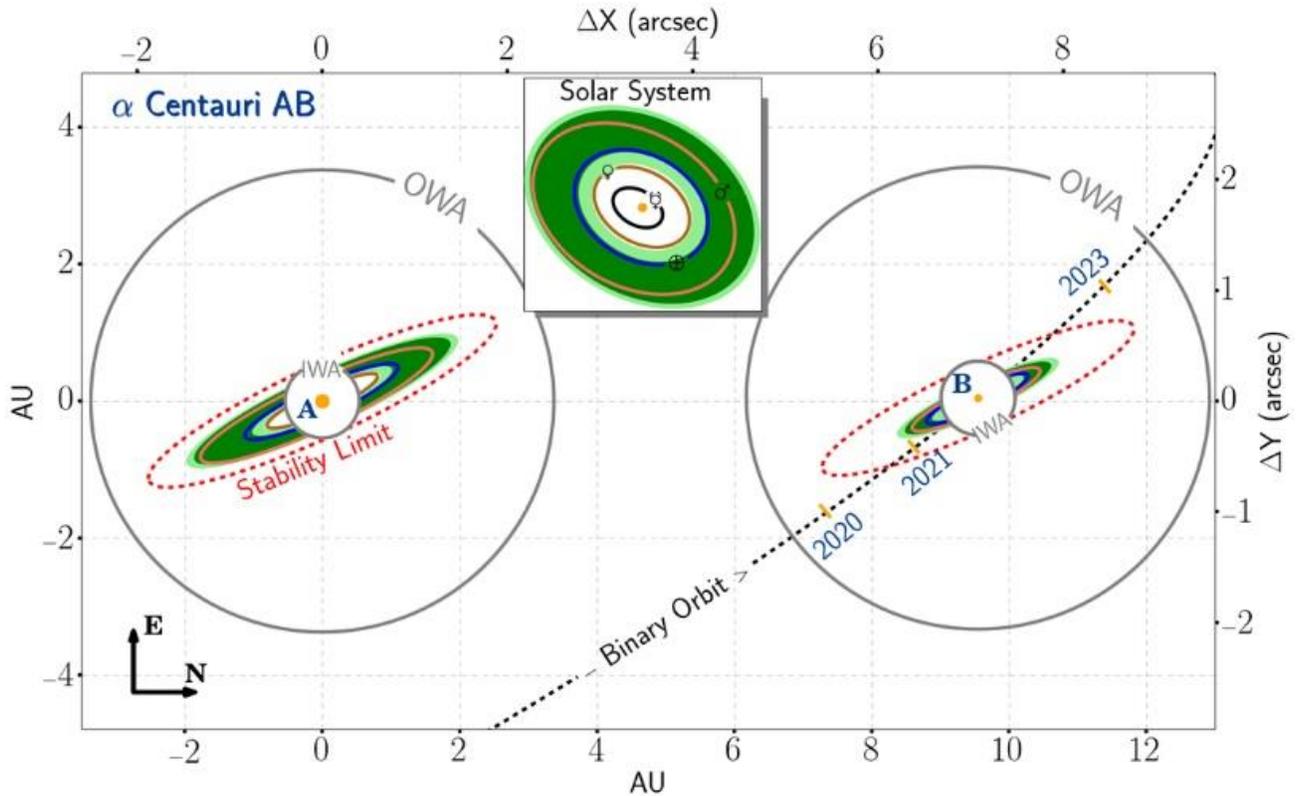

Figure 2. Habitable zones of α Cen A (left) and B (right) in green, along with dynamical stability boundary (red dashed line), 0.4" and 2.5" inner and outer working angles (IWA and OWA). The inset shows the Solar System to scale. Planetary systems of α Cen AB are assumed to be in the plane of the binary (the likeliest scenario) and orbits of hypothetical Venus-like, Earth-like, and Mars-like planets are shown.

Planet formation around the Sun is reasonably well-understood, however the mechanisms (in situ, migration, scattering) for forming planets in multi-star systems remains an open question. We know such planets can form because more than 60 circumstellar (as opposed to circumbinary) exoplanets have been found in binary star systems, including cases dynamically similar to α Cen AB (such as γ Cep), i.e. where the stellar separation is small enough to gravitationally affect the terrestrial planet forming region (<~ 4 AU) around each star. So far all of the planets found in close binaries (~20 AU separation like α Cen AB) are giant planets like Jupiter except for the possible Earth-mass planet announced around α Cen B [13] in a very close 3.2 day orbit The number of such planets is still low, but can be attributed to observational biases (i.e., stellar companions dominate in radial velocity (RV) searches, the small chance of alignment needed for transit photometry, and the incompleteness in searching for stellar companions to known exoplanet host stars).

## 2.2. Confusion with background sources and exozodiacal light

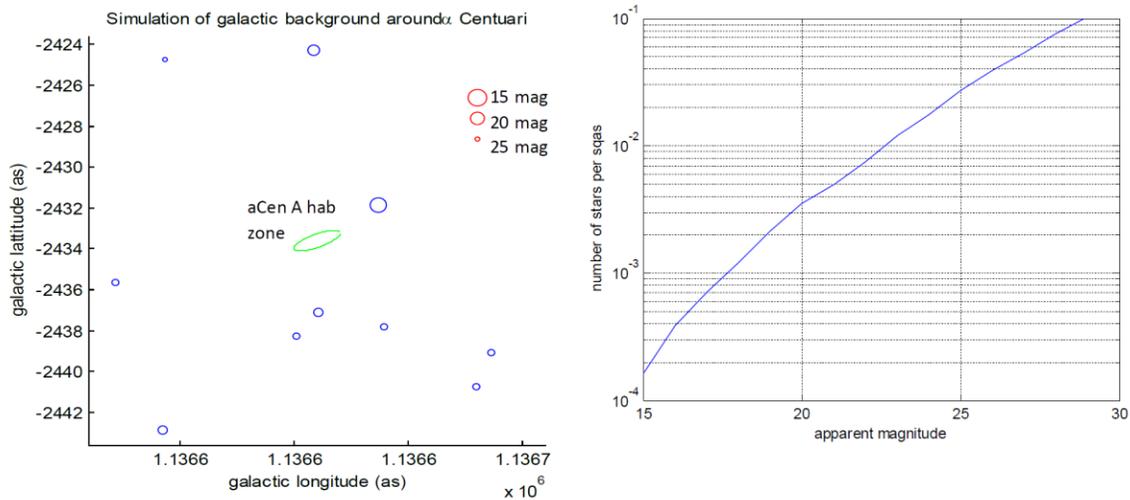

**Figure 3. Simulation of galactic background, showing that chances of confusion are small.**

The analysis in this section is based on the following model:

- A list of simulated stars (by Daniel Huber) using Galaxia code, which implements the Besancon model and simulates different components (thin disk, thick disk, halo, bulge) of the Milky Way [14]
- A simple extinction model of 1.8 mag/kpc on all stars [15]

A sanity check of this model is whether it predicts the expected surface brightness of the Milky Way (21-22mag). It predicts 22mag/sqas, which is very consistent.

Figure 3 (left) shows an ellipse with 1" radius (in green) representing the habitable zone of α Cen A, and simulated background stars. Only stars brighter than an α Cen Earth twin (~25th magnitude) are plotted, qualitatively showing that the chances of confusion are small. This is quantified in figure 2 (right), which shows that there are ~0.03 stars per sqas of apparent magnitude 25 or brighter, so the chances that such a star will be in a potential HZ planet location are less than 3%. Furthermore, these stars are going to be moving at a median proper motion of 0.5mas/year (according to Galaxia data), which is orders of magnitude slower than the expected ~1"/year motion of a planet around α Cen. The proper motion of α Cen itself is ~4" in RA and ~0.5" in Dec, comparable to a planet, which may cause confusion in 2 observations of a planet, but a 3rd observation is sufficient to differentiate proper motion from orbital motion.

*Unresolved galactic background stars*

Figure 3 (right) shows that bright stars contribute more than dim stars to the surface brightness of galaxies. In particular, stars dimmer than 25[th] magnitude together contribute a surface brightness of only 29mag/sqas. In the case of a 30cm telescope and a 0[th] mag star (e.g. α Cen A), this corresponds to contrast of about 1e-12 and is therefore completely negligible if the stars are not resolved. (And if they are resolved, each would be dimmer than 1e-10 contrast and thus not detectable in the first place.)

*Extragalactic background*

Interstellar medium (ISM) extinction is ~1.8mag/kpc, and α Cen happens to lie at an almost 0 galactic lattitude. Therefore, any light from background galaxies has to first travel through tens of kpc of the Milky Way ISM, which ill then almost completely block it.

*Exozodiacal light*

If exozodiacal light is present in the system, it will appear at roughly $10^{-10}Z$ contrast, where Z is the brightness relative to the Solar zodiacal light, or "zodi" (this number is an order of magnitude and varies somewhat depending on telescope size, which star is observed, system inclination, and distance from the star). There is probably at most 100 zodis of exozodical light around both Alpha Centauri A and B [16], so the contribution of exo-zodiacal light will be at most $10^{-8}$ contrast. As will be shown below, our post-processing method can filter out any features at $10^{-8}$ contrast and below that do not move on Keplerian orbits. Thus, as long as exozodi does not have clumps brighter than ~$10^{-10}$ contrast that move on Keplerian orbits, it will be filtered out along with stellar noise. (The possible existence of zodi at $10^{-8}$ contrast also reduces the benefit in designing an instrument capable of raw contrasts better than $10^{-8}$.) If exozodiacal light does form clumps, then this may cause confusion with planets. However, exozodis as dense as 100 zodis are expected to be very smooth, and even if they are not, multi-band imaging orbital phase light curves, and polarimetry are all tools that can help resolve this potential ambiguity.

## 3. STARLIGHT SUPPRESSION STRATEGY AND TECHNOLOGY

Coronagraphic technology to suppress starlight around single-star systems has been steadily advancing and laboratory demonstrations are now better than $10^{-9}$ contrast in broadband light (see, e.g. [17]). However, direct imaging in a binary star system with a small telescope brings two unique challenges: (a) suppressing the light of the second (off-axis) star, and (b) achieving $10^{-10}$ contrast with a coronagraph in a small package and low cost. In what follows, we propose solutions to both (a) and (b). In particular, our solution to (a) relies on a new technique called "Multi-Star Wavefront Control", or MSWC. Our solution to (b) is to relax the raw contrast requirements of the coronagraph to $10^{-8}$, and use a special observing strategy and a new post-processing method called "Orbital Differential Imaging" (or ODI) to reach $10^{-10}$. The strategy relies on spending an extended period of time on the target (months or years), which enables enough images (tens of thousands) to achieve a higher post-processing factor than is possible on large missions that typically only spend a short amount of time on any given target. This allows a relaxation of raw contrast requirements to ~$10^{-8}$, significantly reducing the size, cost, and complexity of the instrument.

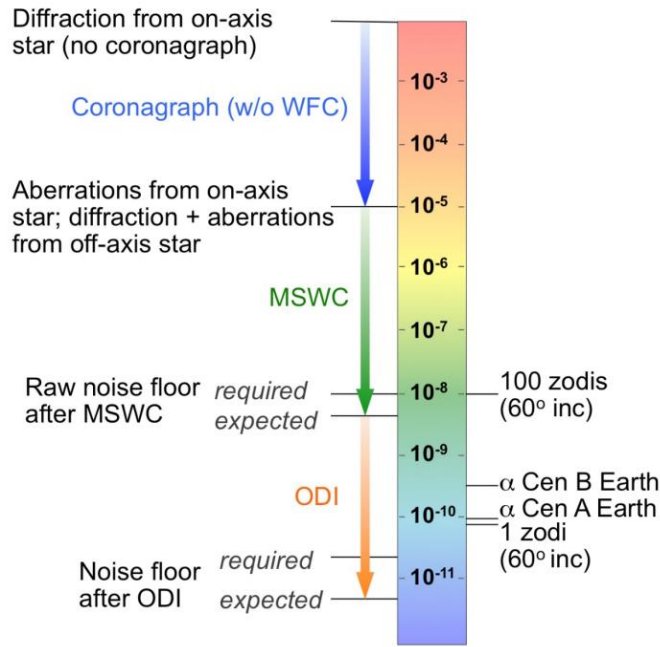

**Figure 4. Approach to achieving high contrast on a small telescope on a multi-star system with modest raw contrast requirements**

The overall approach is diagrammed schematically in Figure 4. Without wavefront control, a single-star coronagraph can typically achieve ~$10^{-5}$ contrast, which is limited by random aberrations in the telescope optics and dynamical low-order errors. Because they are random, they cannot be corrected by a coronagraph and must use active correction. Coincidentally, the diffraction from the second (off-axis) star in the Alpha Centauri would also be roughly at an order of magnitude of $10^{-5}$ contrast for small telescopes (the actual value varies somewhat and depends on which star is observed, size of telescope, and wavelength). Random aberrations due to the second star usually appear at slightly better contrast levels, but still require suppression. Therefore, an active wavefront control system is necessary to suppress the aberrations from both stars, and as we will show below, MSWC can also suppress the diffraction from the second star along with the aberrations. This obviates the need to suppress the diffraction of the second star with a second coronagraph (though baffling the second star can still help if glare or scattered light is an issue). This brings contrast to roughly $10^{-8}$ levels, and then the large number of images enables ODI to brings the contrast to ~$10^{-11}$ levels.

**3.1 Multi-Star Wavefront Control**

The field of high contrast single-star wavefront control is mature with many laboratory and on-sky demonstrations that achieved very high contrasts (see e.g. [17]). Unlike conventional adaptive optics, which typically deals with flattening the wavefront in the pupil plane and achieving a diffraction-limited image, high contrast wavefront control typically deals with removing light in some region of interest (dark zone) directly in the focal plane (which frequently results in substantially non-flat wavefronts in the pupil plane). The typical method is to obtain a measurement of the complex-valued, wavelength-dependent field of the light in the science focal plane region of interest, and then set a deformable mirror in a way that diffracts some light from the star into a pattern that destructively interferes with the unwanted light already present in the region of interest (see e.g. [18]). In order for this to work, the light in the region of interest must be "coherent" with respect to the star, i.e. interfere in amplitude, not in intensity. However, in the presence of two stars, light in the region of interest consists of contributions from both stars, and light from one star is incoherent with respect to the other star. Therefore, any wavefront control system must have the property that it simultaneously and independently creates dark zones for each star. It turns out that this is in fact possible with existing wavefront control systems, simply by applying a special algorithm. We briefly go over the principle here (for more details see [19,20]).

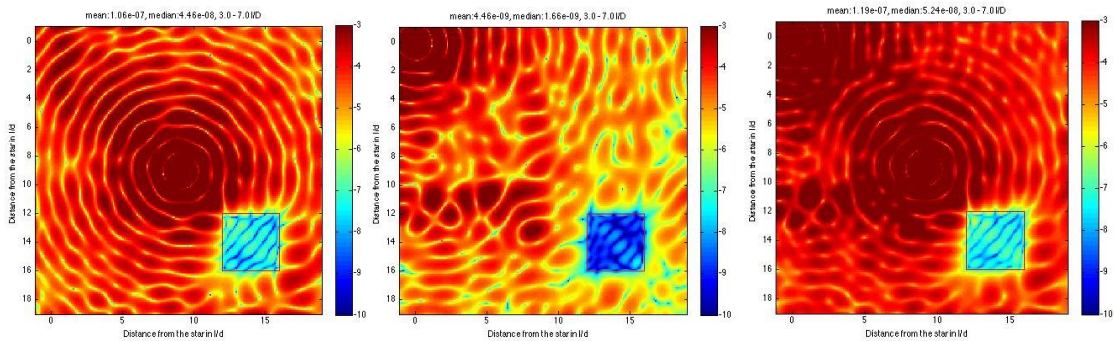

**Figure 5. Preliminary proof-of-principle simulation of Multi-Star Wavefront Control. Left: on-axis star only (in the center); Middle: Off-axis star only (upper left corner); Right: sum of both images. All images have the same deformable mirror setting, showing the creation of a dark zone in the presence of two stars.**

Figure 5 shows a proof-of-principle simulation where for simplicity we assume a system with no coronagraph, in monochromatic light, and with one 32x32 deformable mirror. There are two stars separated by 10 $\lambda/D$ (one in the center and one in the upper left corner), and optical aberrations were added in this simulation, so that the two stars are not just ideal airy patterns. The left image shows the central star with a dark zone region of interest (blue square between 2-6 $\lambda/D$) which to first order is affected by modes on the DM between 2 and 6 $\lambda/D$. The middle image shows only the off-axis star, and the same region of interest appears between 12 and 16 $\lambda/D$ with respect to it. Therefore, to first order, different independent DM modes control the light from the two stars. This enables finding a single DM solution that simultaneously removes light in the region of interest independently for the two stars (left and middle of Figure 5) and thus a combined dark zone is created when the two stars are present together (right of Figure 5). Coupling between modes and the two stars occurs at second order and higher, but a simultaneous solution can still be found as long as the modes are non-degenerate.

Note that this simulation is only a demonstration of the principle that light from two stars can be independently controlled, but is not representative of the full power of this technique, which is still being explored. For example, the size of the dark zone can be made much larger and in theory can be as large as half the dark zone for a single star (to preserve the independent degrees of freedom available on the DM). The presence of a coronagraph or broadband light also does not affect the basic principles of this technique [20].

**Orbital Differential Imaging**

The main feature of Orbital Differential Imaging is that it leverages a large number of images spanning a significant portion of a planet's orbit to extract the planet signal (see Figure 6). This is motivated by two basic principles: (a) signal to noise grows with number of images (at least in the case where the noise is uncorrelated between images); and (b) planets move on Keplerian orbits, so any "speckles" in the image sequence that does not move on a Keplerian orbit can be filtered out (including stationary exozodiacal light). In this section we cover the basic principles, but as shown in a more detailed treatment [21], it appears that at least for some range of assumptions about the nature of noise and systematic errors, factors of 1000 gains in post-processing appear to be possible with this technique (for missions that deliver the requisite long image sequence).

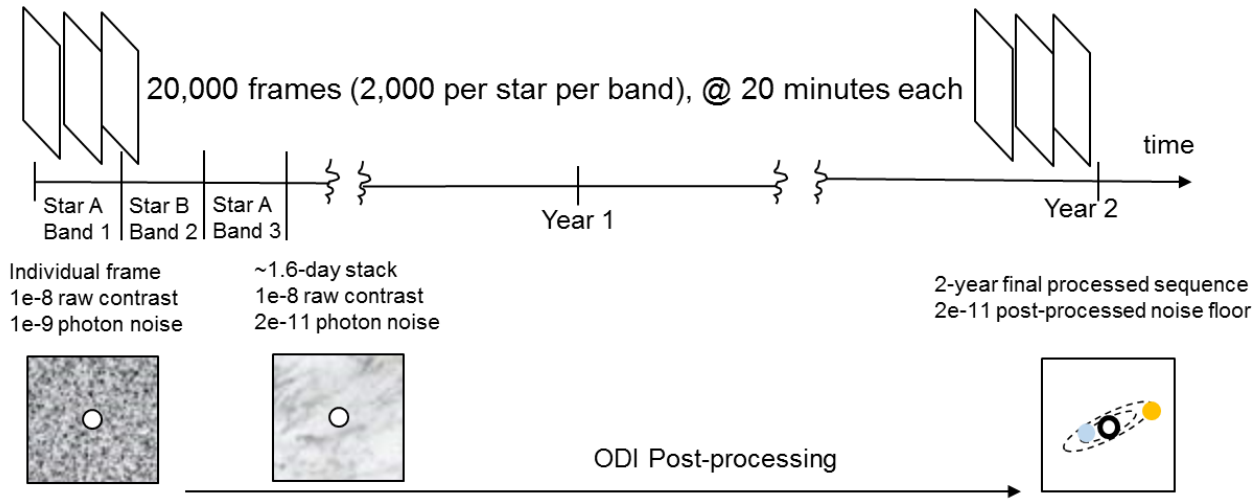

**Figure 6. The proposed post-processing technique (orbital difference imaging) leverages the large number of images collected by spending months or years on a given star to obtain much deeper contrast gain in post-processing than is possible on a single visit.**

Figure 6 shows a simple calculation of how this technique can work on a small telescope. (The numbers were computed for the case of a 45cm telescope and depend on telescope size, bandwidth, system throughput, etc., but the order of magnitude of the numbers remains the same for a wide range of parameters.) 20-minute frames spanning 2 years would result in a total of 20,000 frames (10,000 per star). Assuming a 10% end-to-end system throughput, the photon noise of a $10^{-8}$ contrast speckle will be $\sim 10^{-9}$. A ~2-day stack will reduce the photon noise to $\sim 2 \times 10^{-11}$, so that if the speckle field was perfectly calibrated and known, a 2-day stack of images is sufficient to see a $10^{-10}$ contrast planet at SNR=5. Processing the entire sequence as described in [21] effectively enables the calibration of this field on every 2-day image (though this is treated implicitly in the actual algorithm), revealing planets.

A simulation of ODI is shown in Figure 7. This simulation assumed 5 bands (and other parameters of the mission concept described in the next section), and a post-MSWC raw contrast of $10^{-8}$ at the inner working angle. The top row shows the ODI-processed sequence on Alpha Centauri A where we assumed an Earth-like, Venus-like, and "Pseudo-Mars" planet (at 2.5 AU and Earth-sized). Earth- and Venus- like planets can be seen orbiting the star in the processed sequence. An additional "shift-and-add" technique can be used to further boost signal to noise, where images are shifted and co-added along candidate signal's orbit (Figure 7, bottom). This can recover fainter planets that are not readily apparent on the processed sequence, such as pseudo-Mars.

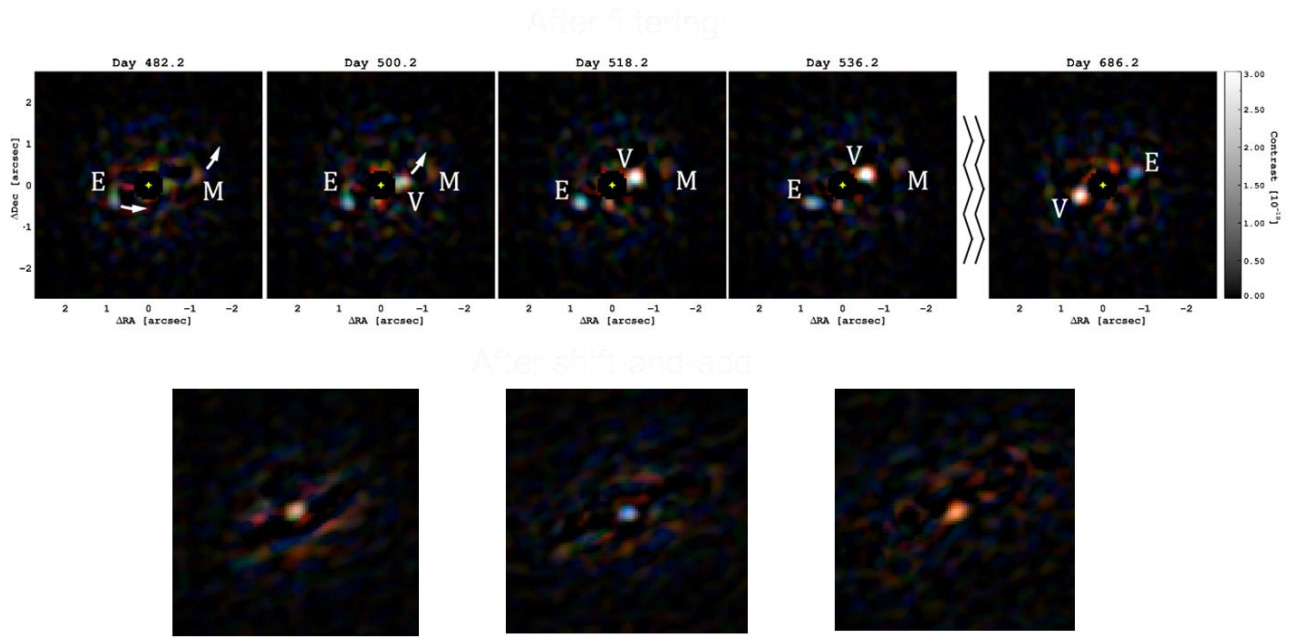

**Figure 7.** Simulation of ODI. Top: image sequence processed by ODI. Bottom: Further improvements can be obtained by aligning and stacking planet images as they orbit the star.

## 4. EXAMPLE OF A MISSION CONCEPT

We have integrated the above ideas and several other innovations together and developed them further into a mission concept called ACESat: Alpha Centauri Exoplanet Satellite. We present a brief overview here, for a detailed description see [22-24].

ACESat is designed to directly image the planetary systems of both Alpha Centauri A and B with the capability of detecting all Earth-like planets and assessing their potential for habitability. In addition, it will detect larger planets as well as planets outside the habitable zones, out to the stability limit of the system at about 2.5 AU, as well as exozodi variations (i.e. exozodi "clumps"). It also has a science enhancement option where a third year is spent observing other stars such as Sirius, Procyon, Altair for larger planets.

It is a small explorer-class (SMEX) 2-year mission consisting of a 45cm coronagraphic telescope launching in 2020. Being a high-performance coronagraph, it has tight mechanical and thermal stability requirements but leverages two factors that make it easier to reach those requirements: (a) it is launching into into an Earth trailing orbit similar to Kepler, and (b) ACESat spends its entire baseline science mission time looking into the same direction, also similarly to Kepler. Both (a) and (b) were important factors enabling Kepler to achieve its great photometric stability. In an analogous way, (a) and (b) help ACESat achieve pointing and thermal stability necessary for high contrast. Another factor aiding stability is the fact that the telescope is fully made out of Silicon Carbide.

The telescope is unobstructed with a PIAA coronagraph architecture (baseline) embedded on the secondary and tertiary telescope mirror. This reduces the number of optics and simplifies the overall design. Such embedding is possible due to the small size of the telescope because it is impractical to have large PIAA optics on the secondary and tertiary of large telescopes. The imager consists of 5 equispaced bands between 400-700nm, split by dichroics which enable differentiation and classification of many planet types.

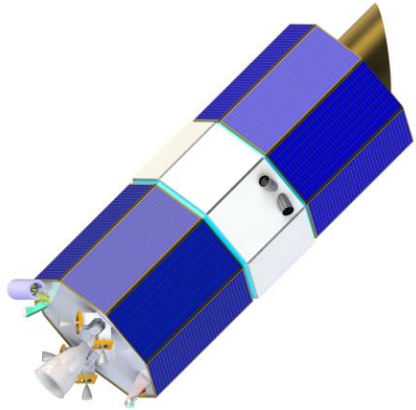

**Figure 8. ACESat: Alpha Centauri Exoplanet Sattellite**

| | |
|---|---|
| Mission Time Life and Orbit | SMEX-Class, launch 2020, 2-Years, Earth trailing |
| Instrument/ Telescope | Unobstructed 45cm, Full Silicon Carbide |
| Coronagraph architecture | Baseline: PIAA Embedded on Secondary and tertiary telescope mirror. |
| Coronagraph performance | $1\times10^{-8}$ raw<br>$6\times10^{-11}$ @ 0.4" (with ODI)   $2\times10^{-11}$ @ 0.7" (with ODI) |
| Wavelength | 400 to 700 nm, 5 bands @ 10% each. |

**Table 2. Top-level parameters of the ACESat mission concept**

## 5.  CONCLUSIONS

We showed that a ~30-45cm telescope (with appropriate high contrast imaging capability) is sufficient to directly image Earth-like and larger planets around α Cen AB. This is enabled by the fact that Alpha Centauri is an extreme outlier in terms of its apparent habitable zone size (~1"). Any other star has a habitable zone that is at least ~3x smaller or contrast levels that are an order of magnitude more challenging. Thus, Alpha Centauri represents an unusual and unique opportunity for direct imaging by small telescopes, potentially enabling a direct image of an Earth-like planet as early as 2020 at low cost.

Potential astrophysical challenges appear to not be an issue. Dynamical simulations show that habitable zone orbits are stable around Alpha Centauri A and B. Chances of confusion with background sources and extragalactic background are low (despite small telescope size and Alpha Centauri line of sight being in the galactic plane) because of the high brightness of the system as well as its large proper motion. Exo-zodiacal light will appear at $10^{-10}$ if it is the same brightness as Solar, and can be removed by Orbital Differential Imaging along with speckles even at its largest expected brightness of 100 zodis.

New starlight suppression technologies promise to solve certain key technological challenges to achieving $10^{10}$ contrast on small telescopes on binary stars. Multi-Star Wavefront Control (MSWC) suppresses the light from both stars without the need of any new hardware development. Orbital Differential Imaging (ODI) leverages a large number of images (for missions that spend months to years on a single target) to achieve a greater that typical post-processing gain and relaxes raw contrast requirements to $10^{-8}$, greatly simplifying the design and cost of the coronagraph for a small telescope.

A 45cm mission concept called ACESat was developed incorporating these concepts, potentially enabling the direct imaging of an Earth-like planet in the early 2020s. It achieves the requires pointing and thermal stability, in part by always pointing in the same direction, and being in a stable Earth-trailing orbit, both of which were important factors to enable the high photometric stability of Kepler.

## 6. ACKNOWLEDGEMENTS

This work was supported in part by the National Aeronautics and Space Administration's Ames Research Center and in part through the development of the ACESat mission concept. The authors gratefully acknowledge the contributions by the entire ACESat team and partners, and thank Billy Quarles for providing Figure 2; Daniel Huber for helping with the simulation in Figure 3; and Elisa Quintana, Jack Lissauer, and Billy Quarles for discussions and references on the stability of the Alpha Centauri system. Any opinions, findings, and conclusions or recommendations expressed in this article are those of the authors and do not necessarily reflect the views of the National Aeronautics and Space Administration.